\RequirePackage{fix-cm}
\documentclass[smallextended,natbib]{svjour3}       % onecolumn (second format)
\smartqed 
\usepackage{graphicx}

\begin{document}

\title{PONDER - A Real time software backend for pulsar and IPS observations at the Ooty Radio Telescope}

\author{Arun Naidu \and Bhal Chandra Joshi \and P.K Manoharan         \and
        M. A. Krishnakumar 
}

\institute{Arun Naidu \at
              National Centre for Radio Astrophysics, Tata Institute of Fundamental Research, Pune 411 007, India \\
              Tel.: +91 020 25719230\\
              \email{arun@ncra.tifr.res.in}           %  \\
             \and
           Bhal Chandra Joshi \at
              National Centre for Radio Astrophysics, Tata Institute of Fundamental Research, Pune 411 007, India \\
              Tel.: +91 020 25719244\\
              \email{bcj@ncra.tifr.res.in}           %  \\
             \and
           P.K Manoharan \at
              Radio Astronomy Centre, National Centre for Radio Astrophysics, Tata Institute of Fundamental Research,
Udhagamandalam (Ooty) 643001, India\\ 
             Tel.: +91 0423 2244962\\
             \email{mano@ncra.tifr.res.in}           %  \\ 
             \and
           M. A. Krishnakumar \at
              Radio Astronomy Centre, National Centre for Radio Astrophysics, Tata Institute of Fundamental Research,
Udhagamandalam (Ooty) 643001, India\\
             Tel.: +91 0423 2244959\\
             \email{kkma@ncra.tifr.res.in}           %  \\ 
}

\date{Accepted by Experimental Astronomy}

\titlerunning{PONDER - Pulsar software backend at the ORT}

 \journalname{Experimental Astronomy}
 \maketitle

\begin{abstract}

This paper describes a new real-time versatile backend, the Pulsar Ooty Radio Telescope New
Digital Efficient Receiver (PONDER), which has been designed to operate along with the
legacy analog system of the Ooty Radio Telescope (ORT). PONDER makes use of the current
state of the art computing hardware, a Graphical Processing Unit (GPU) and sufficiently
large disk storage to support high time resolution real-time data of pulsar observations,
obtained by coherent dedispersion over a bandpass of 16 MHz. Four different modes 
for pulsar observations are implemented in PONDER to provide standard reduced data
products, such as time-stamped integrated profiles and dedispersed time series, allowing
faster avenues to scientific results for a variety of pulsar
studies. Additionally, PONDER also supports general modes of interplanetary
scintillation (IPS) measurements and very long baseline interferometry data recording.
The IPS mode yields a single polarisation correlated time series of solar wind scintillation over a
bandwidth of about four times larger (16 MHz) than that of the legacy system as well as its
fluctuation spectrum with high temporal and frequency resolutions. The key point is that
all the above modes operate in real time. This paper presents the design aspects of
PONDER and 
outlines the design methodology for
future similar backends. It also explains the principal  operations of PONDER,
illustrates its capabilities for a variety of pulsar and IPS observations and
demonstrates its usefulness for a variety of astrophysical studies using the high
sensitivity of the ORT.
\end{abstract}

\section{Introduction}
\label{intro}

\paragraph{} Investigations of the pulsed emission from pulsars, 
which are rapidly rotating highly magnetized compact neutron stars, 
often require very high time resolution time-series data from 
a sensitive radio telescope, such as the Ooty Radio Telescope (ORT).
The pulsed signal is dispersed by free electrons in the 
inter-stellar medium (ISM), causing the pulse to arrive at 
progressively later times with progressively decreasing 
frequencies. The degradation in time resolution and 
the signal-to-noise ratio (SNR) due to this effect
can be compensated by the 
technique of coherent dedispersion \citep{hr75}, 
which is usually computationally
intensive, particularly at low radio frequencies (below 400 
MHz). A new real time software backend, Pulsar Ooty Radio Telescope 
New Digital Efficient Receiver (PONDER),  implementing 
this technique in software using a Graphical Processing Unit (GPU) 
for pulsar observations is described in this paper. The new backend 
also enhances the quality of 
inter-planetary scintillation (IPS) as well as 
incoherently dedispersed pulsar observations by providing 
standard data products in real-time.

\paragraph{} Most pulsar studies require high time 
resolution data. Experiments involving tests of gravitational 
theories 
\citep{tw89,kra98,vbb+01,wt02,lbk+04,ksm+06} and 
the detection of stochastic gravitational wave background using pulsar 
timing arrays \citep{fb90,mhb+13,dfg+13} demand a high degree 
of precision in measurements of the pulsar clock, which requires 
data sampled at fractions of micro-second. High time resolution 
studies of pulsars such as PSRs B0531+21 and B1937+21 
reveal narrow intense highly polarized Giant Pulses (GPs), 
which provide constraints on the location 
and size of emission region as well as the emission 
mechanism \citep{sb95,kt00,jr02,jr03,hkwe03,jkl+04}. 
Observations of microstructure, observed in 
pulsars such as PSRs B1133+16, B0950+08 and J0437$-$4715 
\citep{jak+98,pbc+02,kjv02} place further constraints on 
the emission mechanism for pulsars. 
High time resolution observations also provide high 
precision astrometric measurements as illustrated by the 
distance measurement of PSR J0437$-$4715 system using 
annual-orbital parallax \citep{vbb+01}. 

\paragraph{} Traditionally, specialized hardware using Digital 
Signal Processing (DSP) or Field Programmable Gate 
Array (FPGA) chips were used to implement the coherent 
dedispersion algorithms. In recent years, the availability 
of inexpensive computers has allowed 
implementation of these algorithms in backends, employing 
clusters of computers, e.g. pulsar observing systems at Jodrell Bank, 
Giant Meterwave Radio Telescope, Westerbok, 
Arecibo, Parkes and Green Bank Observatories \citep{jlk+03,jr06,dsd+08,ksv08}. 
A more suitable alternative is now provided by 
GPUs, primarily developed for gaming and with 
several thousand on-board processor cores. These can meet 
the high computational demands of coherent dedispersion, 
particularly at frequencies below 400 MHz, allowing 
on-line coherent dedispersion using a single personal computer (PC). 
PONDER employs a GPU for this purpose.

\paragraph{} PONDER was designed to provide capabilities 
for high time resolution 
observations with the 
ORT, which is an offset parabolic cylindrical antenna, 
used as a sensitive single dish telescope for monitoring pulsars and 
the solar wind. Previous pulsar studies with the ORT were performed with 9 MHz 
of bandwidth with typical time resolution of about 128 $\mu$s. Daytime 
observing at the ORT is allotted to the IPS studies, which are aimed at the
regular monitoring of the solar wind over a wide area of the sky plane
[e.g., \cite{Mano1}]. In the conventional IPS measurements, the
intensity scintillation over a 4 MHz bandwidth obtained with the 
central beam of the correlated-beam system was recorded at a  
sampling interval of 20 ms 
\citep{Mano2}. In addition, it is proposed to use the ORT for 
Very Long Baseline Interferometry (VLBI) observations 
with telescopes in Russia, which also require Nyquist 
sampled voltage data at a very high time resolution.

\paragraph{} The traditional approach for such observations has been high 
speed recording of the data followed by off-line analysis in the 
software due to the large computational resources 
required for analysis, particularly for large bandwidths. 
With the computational power available in PC and GPU boards,  
the routine off-line analysis can be done in real-time. 
This makes available real-time standard data products, 
several orders of magnitude smaller in volume than the raw data    
and simplifies their archiving. 
The software approach also increases the upgradeability and 
flexibility of the backend as new data products can be added in 
future with the  underlying hardware replaced by ever 
improving computational machines available. In addition to the real-time 
coherent dedispersion capability, the design of PONDER was also 
partly motivated by these considerations.

\paragraph{} The organization of the paper is as follows. A brief 
description of the ORT is presented in Section \ref{ort}. 
The description of the hardware (Section \ref{ponderhw}) is 
followed by a discussion of 
the main design considerations 
of PONDER software 
(Section \ref{ponderdescons}) and 
the required software pipelines (Section \ref{pondersw}). 
The results of test observations, demonstrating the 
capabilities of the instrument, are 
presented in Section \ref{ponderill}. Finally, a summary,  
with the possible future development, is 
provided in Section \ref{summary}.

\section{The Ooty Radio Telescope}
\label{ort}

\begin{figure}
\includegraphics[width=1\textwidth]{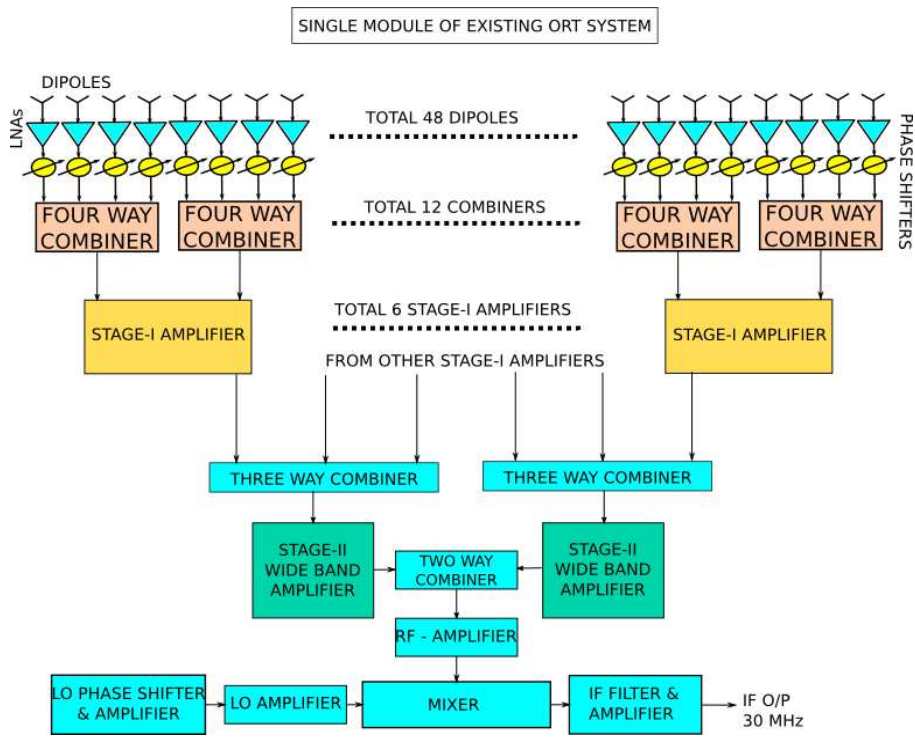}\centering
\caption{Block diagram of single ORT module consisting of 48 dipoles.}
\label{ortfig}
\end{figure}

\paragraph{} The ORT is an offset parabolic cylindrical antenna, 
530 m long in north-south direction and 30 m wide in east-west direction, 
with an effective collecting area of approximately 8500 m$^2$. 
It is erected on a north-south mountain slope with an inclination 
equal to Ooty's geographical latitude ($+11^{\circ}$23$'$), making 
 it an equatorially mounted antenna with its long axis parallel
to the earth's rotation axis \citep{swarup1}. 
 The radio waves reflected by the cylindrical reflector are 
received by an array of 1056 dipoles 
located along the focal line in north-south direction. 
Consequently, the telescope is sensitive to a single linear 
polarisation.  
This array is divided in to 22 sub-arrays, called modules,  with 48 dipoles 
each, which are phased to form module beams. 
These are themselves phased to a given declination,  
using electronic phase-shifters, before combining the 
signals of all modules, allowing the overall beam to be 
steered over a declination range of $-57^\circ$ to $+60^\circ$. 
The telescope beam is steered in the East-West (hour-angle) direction 
by mechanical rotation of the antenna around its long axis. 
The signal from individual dipoles and modules are combined 
by a Christmas Tree network as is illustrated in Fig. \ref{ortfig}.  
Each module output is  mixed with a local oscillator of 296.5 MHz to obtain 
an intermediate frequency (IF) bandpass of 16 MHz centered at 30 
MHz. Different fixed delays are used to synthesize 12 beams 
in the sky. The instrument described in this paper used one 
of the beams of the telescope, i.e., the central beam (Beam 7).
The gain of the antenna is 3.3 K/Jy and the system temperature 
is 150 K.

\section{Design of PONDER}
\label{ponderdes}

\paragraph{} PONDER has been designed to support four main modes: (i) a real 
time pulsar observing mode with filterbank, incoherent 
dedispersion and folded profile data products, (ii) a real time 
coherent dedispersion mode up to a maximum DM of 130 pc~cm$^{-3}$, 
(iii) a real time IPS mode with a bandwidth of 16 MHz and  
(iv) a baseband recorder mode for VLBI observations. The
hardware architecture, design considerations and 
the software architecture of PONDER is described 
in the following sections.

\subsection{Hardware architecture of PONDER}
\label{ponderhw}

\begin{figure}
\includegraphics[width=1\textwidth]{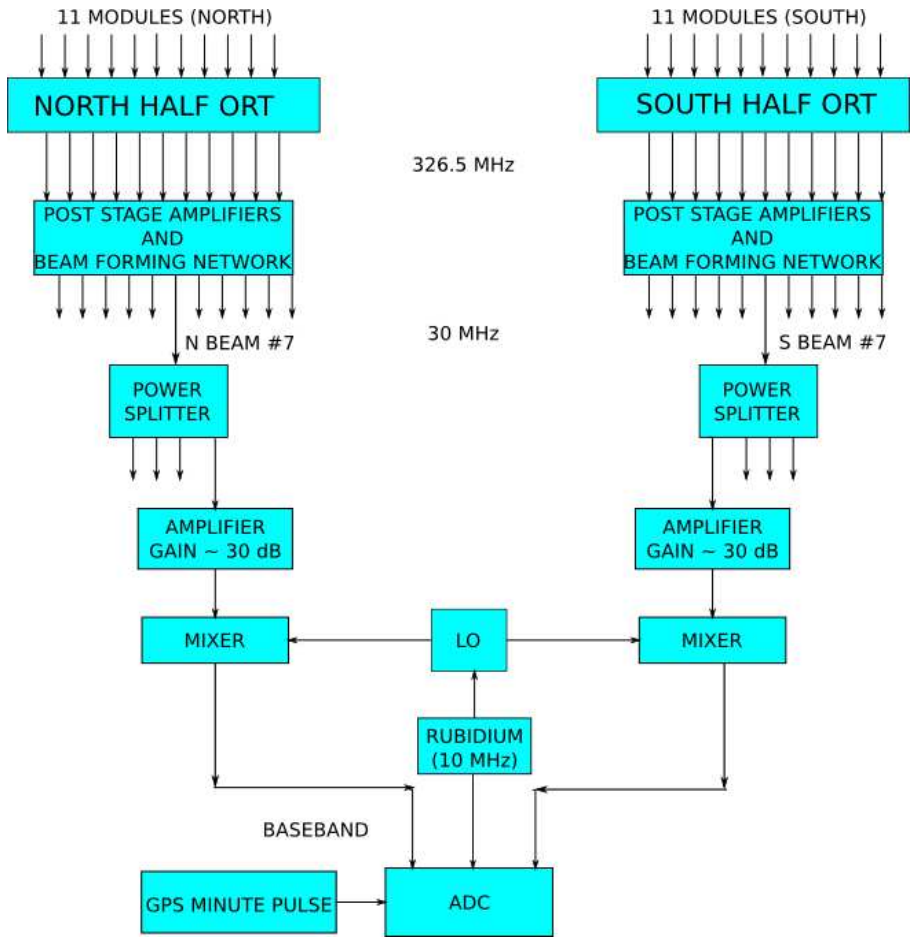}\centering
\caption{Hardware architecture of PONDER.}
\label{ponderhwbldiag}
\end{figure}

\paragraph{} The hardware architecture of PONDER is shown in Fig. 
\ref{ponderhwbldiag}. The 30-MHz IF output of Beam 
7 of each half of the ORT is first amplified by a 30 dB amplifier 
and then mixed with a 38-MHz tone, generated 
using a frequency synthesizer locked to 10-MHz reference 
signal from a Rubidium oscillator disciplined 
by a Global Positioning System (GPS). The down-converted signal 
is recovered with a 16-MHz low pass filter. The power levels 
at the output of the filter can be suitably adjusted with 
variable gain attenuators before digitization using  
analog to digital converter (ADC). The signals from the two halves 
of the ORT are treated identically. 

\paragraph{} The filtered and down-converted outputs of the two halves 
of the ORT are digitized using a two channel Spectrum M3i.2122 ADC board 
mounted on a Peripheral Connect Interface (PCI) slot in 
a Xeon dual processor workstation server.  This card has 
8-bit resolution with values ranging from -128 to 127.
The maximum possible sampling rate the ADC can perform is 250 MHz 
for both the channels and 500 MHz for single channel.
The card has a provision for locking its sampling 
clock to an external reference clock, which was derived 
from the 10-MHz reference of a rubidium clock. The sampling clock 
can be varied from 30 MHz to 200 MHz.
The observatory is equipped with GPS, which 
provides a pulse every minute with the pulse 
edge synchronized to 100 ns accuracy with the Universal 
Coordinated Time (UTC) and this was used to trigger 
the data acquisition by the ADC board. 
Thus, the time of the first sample of the time series is known to 
an accuracy of 100 ns. The ADC has an on-board memory of 1 GB 
to buffer the acquired data, which is streamed to the  Random  
Access Memory (RAM) of the host workstation server without 
any data loss up to 240 million samples per second.

\paragraph{} Based on benchmark tests, carried out on different 
available host PC configurations (as of 2012), 
a low-cost configuration satisfying the requirements for all modes, 
except the coherent dedispersion mode, was selected.
The host used for the digitizer is a server with dual Intel 
Xeon E5645 processors clocked at 2.4 GHz with 6 cores each. 
There is 32 KB of on-chip L1 data cache per core, 256 KB
L2 cache per core, and 12 MB of shared L3 cache. The server 
was equipped with 32 GB of RAM and 9.5 TB of available storage space.
The theoretical peak performance of the system is 
230 Gflops\footnote{http://ark.intel.com/products/48768/Intel-Xeon-Processor-E5645}, 
in single precision. 
Parallel processing can be performed using the 12 cores 
with two way hyperthreading. In the case of PONDER,  the 
hyper threading was disabled because the number of the 
physical cores were more than the number of threads 
required for software implementation. 

\paragraph{}The server is also
equipped with an NVIDIA Tesla K20C with a GK110 GPU, with 
2496 processing cores, arranged in units of 192 streaming 
multiprocessor (SM) and an on-board 5012 MB Graphics Double Data Rate, 
version 5 (GDDR5) memory with a  
bandwidth of 5 GB/s. Each multi-processor has 65536 32-bit 
floating point registers and 32 KB of shared memory. 
The theoretical single precision 
performance of the GPU is 3.5 Tflops and double precision 
performance is 
1.17 Tflops\footnote{http://www.nvidia.com/object/tesla-servers.html}. 
The GPU sits on a full length x16 PCI express slot in the host
machine.

\subsection{Design considerations}
\label{ponderdescons}

\paragraph{} The salient tasks for real time operation are (i) data 
acquisition, (ii) data transfer to host PC-RAM, 
(iii)  Fast Fourier transform (FFT) operation on the data 
to synthesize a digital filterbank, (iv) 
correlation, integration, dedispersion/computation of fluctuation spectra, 
and (v) transfer of processed data from RAM to host hard 
disks.  Nyquist sampling of 16 MHz band from the two halves 
of the ORT  implies a data acquisition speed of 64 M-samples 
per second, which has to be reduced to the final data products 
and recorded in less than a second for a real time capability. 
The  execution time for the required tasks in a serial fashion, 
computed using our benchmark codes, is about 3.8  times the time 
required for data digitization for modes other than 
coherent dedispersion mode and even larger for coherent 
dedispersion.  Hence, to achieve the real time 
computation, parallel computing is required. 

Our tests indicated that the FFT, involved in the digital filterbank 
or coherent dedispersion, take the largest execution time.  
The execution time of a FFT is a function of 
its length, N, which depends on the frequency of observations  
($f_h$), the bandwidth ($\delta f$), sampling time (ts) and the 
dispersion measure\footnote{The dispersion measure is the measure 
of column density of free electrons integrated over the line of sight 
to the pulsar, expressed in units of pc cm$^{-3}$} (DM) and is given by 
the following expression
\begin{equation}
N = \frac{2}{(f_h-\delta f)^2}-\frac{2}{f_h^2} \times \frac{DM}{2.41\times 10^{-4} \times ts}
\label{fftlen}
\end{equation}
\paragraph{}We used both publicly available benchmark 
codes, such as {\it benchfft}\footnote{http://www.fftw.org/benchfft},  
with FFT implemented in publicly available 
Fastest Fourier Transform in the West 
(FFTW) libraries\footnote{http://www.fftw.org/}, as well as a benchmark code 
developed by us\footnote{http://ponderpulsar.sourceforge.net}, % put link for code here
which calculates the execution times separately for 
constant data volume as well as constant number of floating points. 
In particular, the benchmark with constant floating points was developed 
by us to get an estimate of the execution times independent of the 
length of FFT. These codes were also implemented for 
GPU using the GPU programming platform Compute Unified Device 
Architecture\footnote{CUDA is a trademark of NVIDIA Corporations} (CUDA) 
and CUDA C and CUFFT\footnote{http://docs.nvidia.com/cuda/cufft}. 
These benchmarks are useful to explore the effects 
of different cache sizes 
on the execution times of the slowest FFT 
steps in all modes. 

PONDER software was sub-divided into tasks, which were run parallely 
(task parallelism) in different threads. 
The results of our benchmark tests were used in the design of 
the software architecture to balance the computation load with suitable  
task parallelism for all modes of PONDER. 
While task parallelism is adequate for all modes other than the 
coherent dedispersion mode, the computation load for 
this mode is dominated 
by FFT required for dedispersion and is much larger than  
the time required for data digitization.  
For example, observations of a pulsar with a DM of 130 pc~cm$^{-3}$ with 
the proposed backend, the length of FFT is about 135 million points 
(2$^{27}$ points).% can we give an estimate of the execution time 
 Hence, data parallelism (single instruction multiple 
data) amongst multiple cores in GPU was 
used for this mode. The benchmark tests on GPU showed 
that the FFT was performed about 30 times faster than on 
general purpose CPU for 32768 point FFT and the speed-up 
was much larger for longer FFT lengths.

\subsection{Software architecture of PONDER}
\label{pondersw}

\paragraph{} 
Broadly, the software architecture 
is divided in two main processes a) ADC process (process I) and b) 
data reduction and recording process (process II). The Process I
is run as root and process II runs as user. The real time 
processing requirement of the receiver are met by 
inter-process communication(IPC) implemented with shared memory and 
task parallelism using POSIX threads. 
These are implemented using the {\it pthread} library, which is 
available on all modern UNIX systems. The communication 
between the two main processes is achieved using shared 
memory and  is protected by another IPC 
device, namely the semaphores, to achieve synchronization 
between the processes. Mutexs are used for synchronization 
between threads. All the FFT operations, implemented on the 
host server workstation with dual Intel Xeon E5645 processors, 
used the FFTW
library. The receiver also uses CUDA API developed by NVIDIA.  
The PONDER software   
implements the four primary modes of operation of the 
backend as mentioned in the beginning of Section 
\ref{ponderdes}. The structure of the software and 
its operation for each of the major modes of operation 
of PONDER are described in detail in the following sections.

\subsection{Real-time pulsar mode with incoherent dedispersion}
\label{rpsric}

\paragraph{} In this mode, PONDER is used for pulsar observations with 
incoherent dedispersion. This is the simplest way to compensate 
for the effects of pulse dispersion due to the ISM. This method 
involves splitting the incoming frequency band into narrow 
channels. The voltage  in each channel is then converted to power 
and consequently the phase information is lost. The signal 
in each channel therefore suffers from a residual dispersive delay 
given by
\begin{equation}
\Delta t_{res}~~ \simeq ~8.3~ \times 10^6 \frac{\Delta f_{ch} \times DM}{f_{ch}^{3}}\,\,\,(ms)
\label{tres}
\end{equation}
where, f$_{ch}$ is the frequency in MHz corresponding to the channel,
$\Delta$ f$_{ch}$ is the bandwidth in MHz per channel. The effective 
time resolution of this mode therefore depends on DM (in pc-cm$^{-3}$) and the 
bandwidth per channel and is worse than that achievable with  
coherent dedispersion. For large DM pulsars, the limit on time
resolution is in any case set by scatter-broadening, which increases 
approximately with the second power of DM \citep{Romani}. 
The scatter-broadening for these pulsars is much 
larger (typically greater than 4 ms) than the corresponding dispersion 
smear for 
1024 channels across 16 MHz bandpass (
3.8 $\mu$s/DM pc-cm$^{-3}$). For such pulsars, the excessive computation required 
by the coherent dedispersion mode is not necessary and the incoherent 
dedispersion is sufficient. 

\begin{figure}
\includegraphics[width=1.0\textwidth]{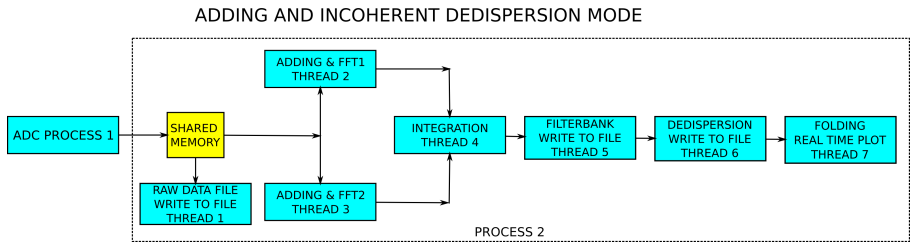}\centering
\caption{Flow chart showing the Adding in Phase and Incoherent
  dedispersion mode (AID). All the tasks in process 2, indicated by 
  a dashed box, are implemented in concurrently running threads 
  labeled by thread number for reference in the text. 
The threads and processes are shown in blue color and the shared memory
in yellow. }
\label{flowdiagram1} 
\end{figure}

\begin{figure}
\includegraphics[width=1.0\textwidth]{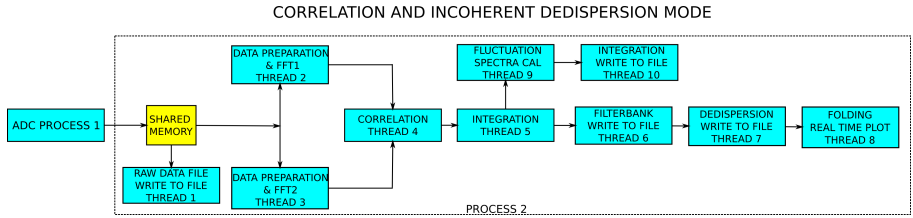}\centering
\caption{Flow chart for the correlation and Incoherent dedispersion 
mode (CID)and IPS mode. All the tasks in process 2, indicated by 
  a dashed box, are implemented in concurrently running threads  
  labeled by thread number for reference in the text. 
The threads and processes are shown in blue color and the shared memory
in yellow. }
\label{flowdiagram2} 
\end{figure}

In the incoherent dedispersion mode, appropriate time
delays, given in Eq. \ref{dm}, are applied to the detected 
signal  in each 
channel to compensate for the dispersive delay across channels.
\begin{equation}
\Delta t_{DM}~~ \simeq ~4.15~ \times 10^6 ~ (f_{ref}^{-2}~-~f_{chan}^{-2})~ \times DM \,\,\,(ms)
\label{dm}
\end{equation}
where f$_{ref}$ is the reference frequency of the bandpass of the 
receiver and f$_{chan}$ is the frequency 
corresponding to channel under consideration (both frequencies 
expressed in MHz) . In this mode, there are two different sub-modes 
to select: (i) Adding Incoherent Dedispersion (AID) mode and  (ii) 
Correlation Incoherent Dedispersion (CID) mode. In the former, 
the digitized data from the two halves of the ORT are 
added in phase before dedispersion, whereas in CID mode, the 
data from both the channels are correlated before performing 
dedispersion. While AID mode provides $\sqrt 2$ better sensitivity 
than CID mode, the latter is less prone to radio frequency 
interference (RFI) as the RFI picked in one half of the ORT 
will in general not be correlated with that in the other 
half. 
Figures \ref{flowdiagram1} and \ref{flowdiagram2}
show how these two sub-modes are 
implemented. All the threads are shown in blue 
and the shared memory is shown in yellow.

\subsubsection{Adding and Incoherent dedispersion mode(AID)}
\label{rpsricaid}

\paragraph{} In this mode, all the threads shown 
in the Fig. \ref{flowdiagram1}, are executed. The 
pipeline was designed with different tasks distributed 
over the minimum required number of threads to achieve a 
balanced computation load. Each process/thread 
is executed on a different core/processor of the host PC. 
The data from the process I are passed on to two threads 
(threads 2 and 3) of process II. Successive blocks of 128 MB 
are directed to threads 2 and 3 alternately by process I. These 
are also passed to thread 1 for a record of raw data.  In the
threads 2 and 3, the data are added in phase and FFT is performed 
using the FFTW library. Two concurrent threads were used for 
addition and FFT as the compute to observed ratio 
(COR)\footnote{the ratio of the execution time of a thread to the 
data digitization time, which should be less than 1 for real-time capability}
for a single thread exceeded 1. The resultant spectra are passed onto
the next thread (4), where each spectra is squared to get 
power and added to achieve the desired temporal resolution. 
These power spectra are passed on to the 
filterbank thread (5), where they are converted into 
SIGPROC\footnote{www.sigproc.sourceforge.net} 
filterbank format. The filterbank thread writes the data in 
this format to the hard-drive and forwards it to the 
dedispersion thread (6), where it is incoherently dedispersed to 
a user specified DM and written to a file as binary data 
with an appropriate SIGPROC header. The 
dedispersed time series from this thread is folded with 
appropriate period, obtained from the TEMPO2 predictors 
\citep{tempo2} by the fold thread (7). The folded profile 
is generated  in real-time and written periodically to disk 
and is displayed on the screen by the Graphical User Interface 
(GUI). The final average profile produced by the fold 
thread is written as an ASCII file. While the output of the 
dedispersion and fold threads is compatible with SIGPROC 
time series and profile format, the code for these threads  
was developed independently by us. The tasks are distributed 
to the seven threads so that the execution time for each is less than 
the data digitization time (COR $<$ 1.0).

The filterbank data, 
dedispersed time series and folded profiles are written 
with SIGPROC headers \citep{lor01b}, which contain essential 
information about the observations such as the modified Julian date 
(MJD) of observations, the frequency of observations, 
bandwidth per channel, total duration of observations, 
nominal DM and period. The software allows either a simultaneous 
recording of raw data, filterbank data, dedispersed time series 
and folded profiles or a recording of selected data products 
out of these based on the parameters supplied by the user 
through GUI.

\subsubsection{Correlation and Incoherent dedispersion mode (CID)}
\label{rpsriccid}

\paragraph{} In this mode, all threads in process II 
(Fig. \ref{flowdiagram2}) other than threads  
9 and 10 (required for IPS mode) are used. The data passed on by 
process I through the shared memory are first sorted into 
two individual channels representing the two halves of 
the ORT in threads 2 and 3. Then, the FFT is performed 
on the data from individual halves by these threads.  
Two concurrent threads are required  for these operations 
as the COR for a single thread exceeded 1. 
The spectra of North and South halves, obtained respectively 
by threads 2 and 3, are then correlated by thread 4 by 
multiplication of the North and South spectra. The correlated 
spectra are accumulated up to user specified 
integration time by integration thread (5). The rest 
of the operation of this mode is similar to that of 
AID mode and is handled in a similar fashion by threads 
6 to 8. The tasks were distributed 
to the eight threads used  so that each has a COR much less than 1.

\begin{figure}
\includegraphics[width=0.75\textwidth]{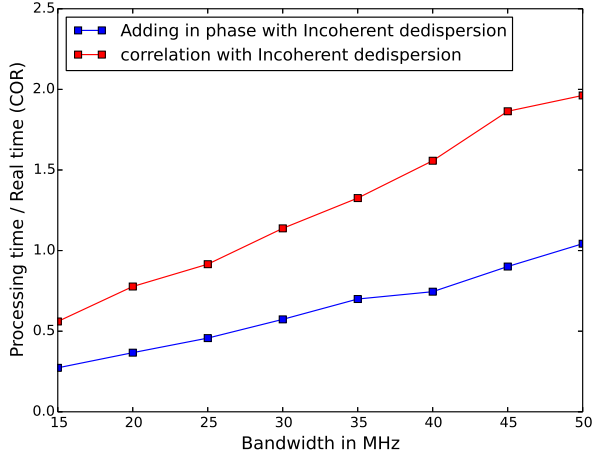}\centering
\caption{The ratio of processing time to real time 
for AID and CID modes. The mode can run in real time 
if the value is less than 0.9.}
\label{freqbench}    
\end{figure}

\paragraph{} Both the above modes perform incoherent 
dedispersion. The performance of incoherent dedispersion 
program as a function of bandwidth is plotted in  Fig.  
\ref{freqbench}. As the sampling frequency is increased, 
the ratio of processing time to real time increases. 
Since CID is more computationally intensive, 
the increase is much steeper. The COR for AID is less than 1 until 45
MHz, while CID can operate until 25 MHz with a COR of less than 
unity. For the current 16 MHz system, both modes operate in real-time. 
This shows the longevity of the receiver if the bandwidth 
is increased in future upgrade to ORT.

\subsection{Inter planetary scintillation mode (IPSM)}
\label{rips}

\paragraph{} The flexibility of the data acquisition of 
the PONDER system has motivated us to initiate
an interplanetary scintillation mode, which provides
observation over a wider bandwidth (16 MHz) 
than the legacy system currently in 
use at the ORT. In this mode, threads 2 to 5 and 9 to 10  of process 
II ( Fig. \ref{flowdiagram2}) are used. 
The data from the process I are passed on to threads  
2 and 3 of process II. The processing for IPS is 
similar to CID mode of pulsar observations and is carried 
out by threads 2 to 5, 
where the data from the two halves of the ORT are correlated after 
performing FFT and the cross spectra thus obtained are 
integrated up to a user specified integration time 
(typically $\sim$ 1 ms). 
The intensity scintillation time series obtained from the
PONDER can support a time resolution starting from 64 $\mu$s 
or higher. 
The integrated spectra are collapsed and the resultant time-series is 
written to hard disk by thread 9 for offline analysis. 

The data processing of the IPS signal primarily involves
obtaining the temporal power spectrum of the intensity
collapsed scintillation time series. The temporal spectrum 
of a given scintillating radio source is computed for the 
required frequency range and resolution, i.e., by
appropriately choosing the sampling rate and the 
length of the data. In the real-time operation, the 
collapsed scintillation time series is further integrated 
to the required time resolution followed by computation 
of its fluctuation spectra over appropriate length of 
data by thread 10 by performing suitable length FFT. 
These spectra can further be  accumulated to improve 
signal-to-noise ratio.  
The accumulated spectra are written as final data product 
by this thread. 

\subsection{VLBI/raw data  mode of PONDER}
\label{vlbi}

\paragraph{} PONDER can be used as a baseband data recorder. This mode 
of operation is primarily useful for VLBI observations, but is 
also useful for developing an offline pipeline for any 
astrophysical investigation not covered by the current 
functionality of PONDER. In this mode, only the process I 
and thread 1 of
process II are used (see Fig. \ref{flowdiagram1}). The 
baseband input from the two 
halves of the telescope are digitized by the ADC in 
process I. Each 128 MB block of data contains samples from 
both channels arranged in an interleaved fashion. This block is
transferred to thread 1 of process II . This thread 
can be configured to write the raw data as it is 
to a file on hard disk or to convert it into standard 
VLBI format. The VLBI format currently implemented 
is Mark 5B\footnote{http://www.haystack.edu/tech/vlbi/mark5/}. 
 
\subsection{Phase-coherent dedispersion mode for pulsar observations (PCD)}
\label{rpsrpcd}

Phase-coherent dedispersion completely removes the dispersive 
effects of ISM, which can be described as a cold, tenuous plasma.  
The frequency response function  
resembles a unity-gain phase delay filter (Eq. \ref{filter})  
and its inverse is used to deconvolve the 
observed signal \citep{Hankins,Hankins1}. 

\begin{equation}
H(f ~ + ~ f_0)~=~exp\left(i\frac{8.3 \times 10^3 \times \pi DM f^2}{f^2_0 (f + f_0)}\right) 
\label{filter}
\end{equation}

Here f$_0$ is the center frequency of the observed band and all 
frequencies are expressed in MHz. DM is given in pc-cm$^{-3}$. 

While the required filtering operation can be done as a 
convolution in the time domain, it is more efficient to 
perform this in the 
frequency domain, where the observed signal is simply 
multiplied with the discrete form of inverse of the 
frequency response function (H$^{-1}$) along-with an  appropriate 
taper function, which is required  to take care of sample to 
sample leakage in an FFT with square window. 
The length of the impulse response of the filter in 
Eq. \ref{filter} has to be larger than the dispersion 
smear and has to also take into account the edge effects 
in a cyclical convolution. 
For a typical DM of 130 pc~cm$^{-3}$ at the ORT 
observing frequency of 326.5 MHz, this length  is about 
2$^{27}$. While such long FFTs are difficult to execute on the 
host server with Xeon processor for 16 MHz bandwidth, these 
can be easily implemented on a GPU using CUDA C.

\begin{figure}
\includegraphics[width=0.9\textwidth]{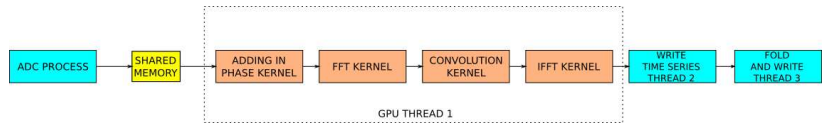}\centering
\caption{Signal flow diagram showing coherent dedispersion. 
All the blocks in orange are kernels. The GPU thread in process 
II is indicated by dotted box around the kernels.}
\label{gpuflow}
\end{figure}

\paragraph{} 
The CUDA C program consists of both host (CPU)
and device (GPU) code. So a traditional C compiler will not
accept the code. The code need to be compiled by a compiler 
that recognizes and understands both host and device code.
We used CUDA C compiler by NVIDIA called NVIDIA C COMPILER (NVCC).
NVCC processes a CUDA program using CUDA keywords to separate 
the host code with the device code. The device code is marked with 
CUDA keywords for labeling data-parallel functions, called kernels,
and their associated data structures. Each CUDA kernel can execute a massive 
number of threads. All these threads run in parallel. 
The signal flow diagram for the implementation of coherent dedispersion 
on GPU is shown in the Fig. \ref{gpuflow}.
The data stored in the host are sent to the GPU device (global memory) 
using shared memory. The GPU process consists of four kernels. 
The Adding in phase (AIP) kernel consists of 2$^{27}$ threads, 
where the interleaved data from both the halves of ORT 
are added in phase. After the 
AIP kernel, the data are passed through the FFT kernel, which  
performs the FFT using the CUFFT API. The resultant spectra are multiplied 
with the filter function $H^{-1}$ by the convolution kernel. 
In GPU programming, the execution time is governed by the number of 
global memory accesses (GMA) performed by each thread. 
In the convolution stage, each thread requires corresponding 
FFT output value and corresponding chirp function value.
The FFT output value is stored in the register memory of each 
thread and the chirp function value is calculated by each thread 
every time the kernel is executed, instead of storing the 
values in the global memory to maintain the GMA to 1. 
Finally, the filtered signal is converted to time domain in 
the IFFT kernel after applying the taper function. 
All the kernels are run in sequence and output of the IFFT kernel 
is sent to data output thread, where the dedispersed time 
series  is written to the hard-drive using SIGPROC format. The 
last thread folds the dedispersed time series to the required 
number of bins and writes the folded profile to the hard-drive. 
The three threads in the process II run concurrently.

\begin{figure}
\includegraphics[width=0.75\textwidth]{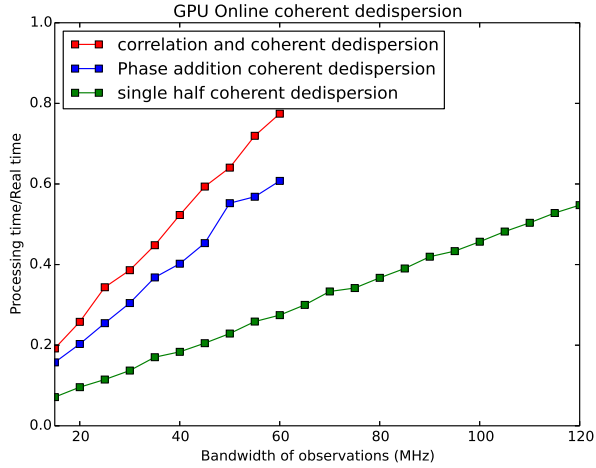}\centering
\caption{The ratio of processing time to real time as a 
function of bandwidth for different modes of coherent 
dedispersion at maximum possible DM . 
The mode can run in real time if the value is less than 0.9.
$N$ is 2$^{27}$ which determines the maximum DM. The red and 
blue curve show the COR for processing data from both halves 
of the ORT, whereas the green curve shows the COR for one 
half only.}
\label{gpucoherent}
\end{figure}

\paragraph{} The performance of PONDER in PCD mode is shown in Fig. 
\ref{gpucoherent}. The  processing time to real time ratio 
is well below unity for the design bandwidth of 16 MHz, 
which shows that PONDER can routinely carry out 
coherent dedispersion at the ORT. The current limitation on 
$N$ is due to the on board memory of 5 GB. In this
case, it translates to 2$^{27}$ points. With this limitation in 
mind, the maximum DM  
that can be observed depends on the bandwidth 
and frequency of observation. The plots shown in Fig. \ref{gpucoherent}
are independent of observing frequency. The maximum possible 
observable DM should be calculated from the observing frequency 
and bandwidth using Eq. \ref{tres}.
In the case of PONDER operating at 326.5 and 16 MHz bandwidth, this DM  
is 130 pc~cm$^{-3}$.
In actual practice, 
particularly for low latitude pulsars and pulsars towards 
the Galactic center, the time resolution is limited by the 
scatter-broadening of the pulse at 326.5 MHz to a much 
smaller DM value, as discussed in Section \ref{rpsric}, 
and coherent dedispersion is not necessary except for 
studies of GPs and scatter-broadening by the ISM. 

 \subsection{Graphical User Interface (GUI)}
\label{gui}

A Graphical User Interface (GUI) was developed, using Perl-Tk package,
for user-friendly observations with PONDER. The GUI allows the user to
customize and start the observations as per the user inputs 
and select the possible data products, listed in Table \ref{tabprod}, for 
the desired mode. Three different types of observations can be customized 
by the user - (1) Individual source (pulsar) mode, (2) List mode and (3) 
IPS mode.

\begin{table}
\caption{Data products available for different modes}
\label{tabprod}
\begin{tabular}{cccccc}
\hline\noalign{\smallskip}
Mode & Rawdata& filterbank& dedispersed& real time  & fluctuation   \\
     &  or VLBI & data &  data & folded profile & spectra \\
\noalign{\smallskip}\hline\noalign{\smallskip}
AID      & yes              &  yes            &  yes             &
yes  &  no                   \\
CID      & yes              &  yes            &  yes             &
yes    & no                 \\
PCD      & yes              &  no             &  yes             &
 yes  & no                    \\
IPS      & yes              & no              &  no              &
no    &  yes  \\
\noalign{\smallskip}\hline
\end{tabular}
\end{table}

\section{Illustration of the capabilities of PONDER}
\label{ponderill}

\paragraph{} PONDER was tested for different types of 
pulsar and solar wind observations after its implementation. 
A brief account of these test astronomical observations 
illustrating the capabilities of the instrument are 
highlighted in this section.

\begin{figure}
\includegraphics[width=1\textwidth]{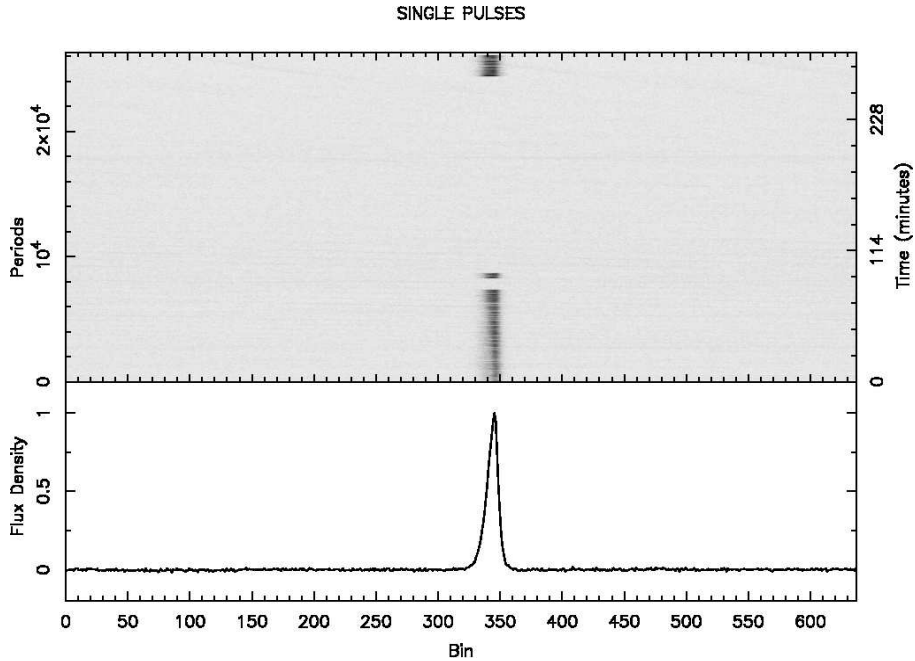}\centering
\caption{The top panel shows the gray scale single 
pulse plot of PSR J1709$-$1640 showing the 
long null observed using PONDER. The bottom panel is the 
corresponding integrated profile. The x-axis is in bins and the
y-axis on left is in periods and on the right is in minutes.
The arrival of the pulse at a constant phase (bin) indicates 
the stability of the system for long observations.}
\label{sp1709} 
\end{figure}

\paragraph{} During tests, PONDER provided high quality time 
series on pulsars. The variation in single pulse 
energies for strong pulsars, such as PSR B0329+54, was useful 
to characterize saturation effects in the backend and devise
strategies to alleviate such effects, particularly for strong 
single pulses, such as Giant pulses. With 8-bit digitization,
the receiver also has high dynamic range, useful for fluctuation 
studies as well as  modulation index studies.

\paragraph{} Fig. \ref{sp1709} shows a single pulse plot of 26265 pulses 
of PSR J1709$-$1640 (P$=$0.653 s, DM$=$24.87 pc\,cm$^{-3}$) 
observed over 4.8 hours, illustrating the long term stability 
of PONDER, useful for long monitoring 
observations of such pulsars with interesting single pulse behaviour.
A clear cessation of emission for more than 2 hours is seen, 
which suggests prominent long nulls in this pulsar. 
With its capability of real time dedispersion, even at high time resolution, 
such long observations do not require large storage space and 
can be analyzed quickly allowing studies of pulsars with 
interesting behaviour. In addition, these real-time single 
pulse plots, together with integrated profiles, can be used to 
assess data quality, while the simultaneous recording of raw data 
in SIGPROC filterbank files helps in a more refined off-line analysis.

\begin{figure}
\includegraphics[width=1\textwidth]{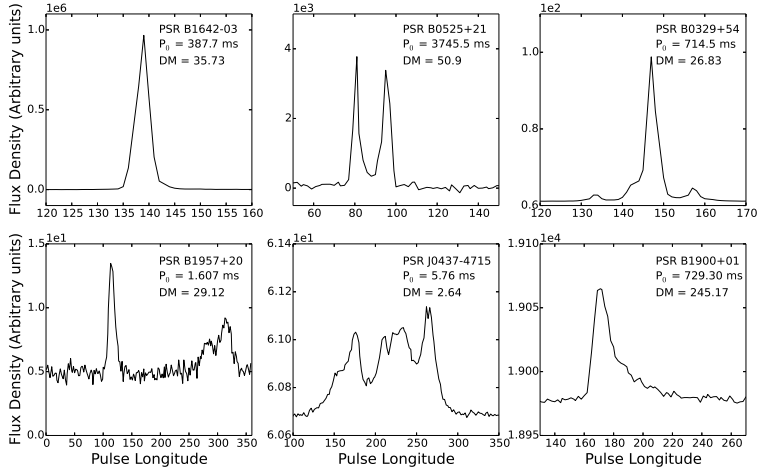}\centering
\caption{This figure shows the integrated 
profiles of a few pulsars, with a range of periods and 
DMs, observed during the test runs 
with PONDER. The flux density in arbitrary units is plotted against 
pulse longitude in degrees (one period $=$ 360$^0$ degrees), 
for appropriately selected longitude range, where the pulsed 
emission is seen, in 
each panel. The pulsar name, period and
DM are given in the top left corner of each panel.}
\label{multip}
\end{figure}

\begin{figure}
\includegraphics[width=1\textwidth]{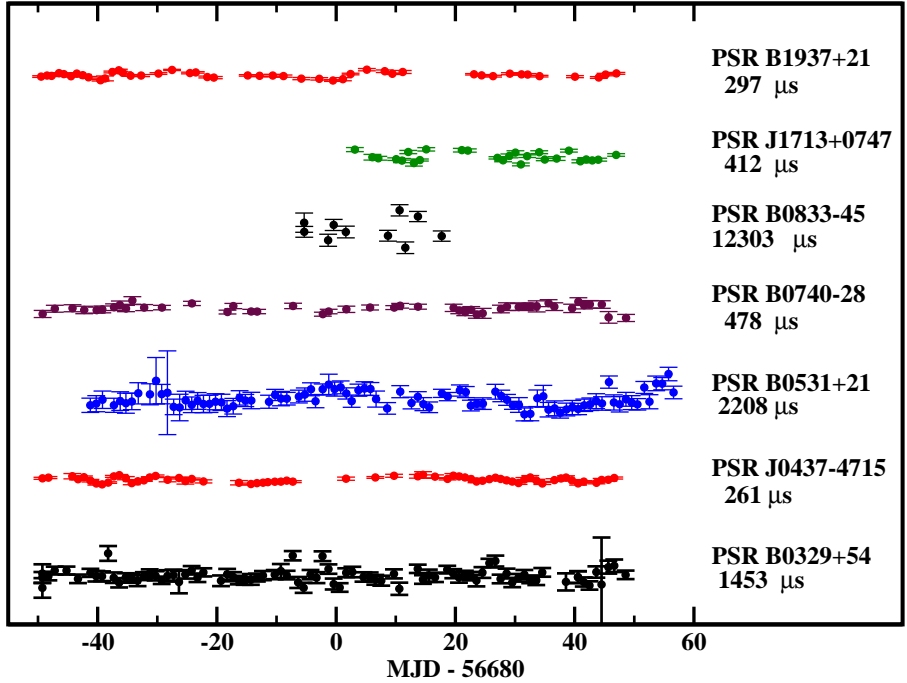}\centering
\caption{The timing residuals for 7 pulsars observed over 100 days. 
These residuals were obtained after subtracting barycentre corrected 
times-of-arrival of the pulses from those predicted using the known 
rotational model of each pulsar. The name of pulsar and peak-to-peak 
amplitude of the residuals is indicated on the right hand side of 
the plot. }
\label{timing} 
\end{figure}

\paragraph{}  Radio pulsars, with periods ranging from 1.57 ms 
to 3.74 s and DMs ranging from 2.64 to 439 pc\,cm$^{-3}$, have been 
observed in order to test PONDER. A selection of integrated profiles 
from these test observations is shown in Fig. \ref{multip} and 
illustrates the variety of average emission studies that can 
be carried out using PONDER. Multi-epoch monitoring with PONDER 
of such profiles can be useful for pulsar timing and mode-changing 
studies. Repeatability of these profiles over several days is 
also a useful test of the stability of the backend. 
In the validation phase of PONDER, a set of pulsars was 
observed in frequent test observations. As these were test 
observations, the time-series and profiles were recorded with 
a coarse sampling time (about 64 $\mu$s or more) in order to avoid 
excessive data volume. The repeatability of profiles 
is illustrated by the timing residuals for 7 pulsars, obtained using 
timing analysis software TEMPO2, from such multi-epoch 
test observations (Fig. \ref{timing}). 
While these test data are useful for validating 
the repeatability of profiles,  it must be noted that 
these tests were not intended as high precision timing 
observations. Nevertheless, peak to peak variation in 
these residuals range from 260 $\mu$s to 12 ms 
(root-mean-square residuals of about 50 $\mu$s) over 
about 100 days demonstrating the stability of instrument 
as well as its capability for routine pulsar timing observations 
for large pulsar timing experiments. In case of PSR J0437$-$4715, 
relatively larger residuals, despite its high SNR profile, are probably due to 
profile changes that were observed. We believe that these could be 
due to Faraday rotation and the single polarisation nature of 
the data.
Lastly, observations 
of a large sample of pulsars at low frequencies can provide 
estimates of scatter-broadening in ISM and results of such 
a study are described in a separate work \citep{k15}. 

\begin{figure}
\includegraphics[width=1\textwidth]{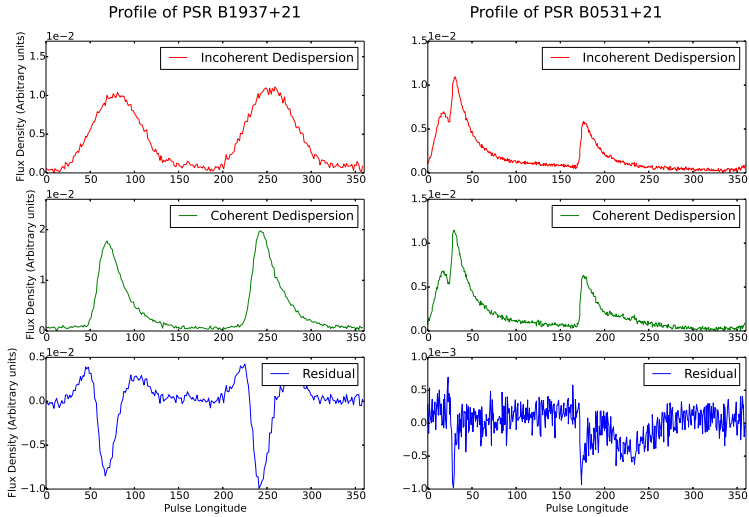}\centering
\caption{Comparison of profiles for pulsars PSR B1937+21 (left panels) 
and PSR B0531+21 (right panels) obtained after  incoherent and 
coherent dedispersion.  For each pulsar, the top panel shows 
the incoherently dedispersed integrated profile, the  
middle panel shows the coherently dedispersed integrated profile and 
the bottom panel shows the difference between the profiles in the 
top and middle panels. The flux density in arbitrary units is plotted 
against the pulse longitude in degrees in each panel.}
\label{MSPIP}
\end{figure}

\paragraph{}The fine structure of pulse emission can be better understood 
using high time resolution data on a radio pulsar, which 
requires coherent dedispersion. As this was the main motivation 
for developing this system, we carried out high time resolution 
observations of several pulsars with real-time coherent dedispersion 
being performed on the GPU. The integrated profiles of two fast 
MSPs, PSRs B1937+21 (P = 1.5 ms, DM = 70 pc\,cm$^{-3}$) 
and B0531+21 (P = 33 ms, DM = 56.8 pc\,cm$^{-3}$), with a phase 
resolution integrated  up to 1 $\mu$s from the base 
resolution of 62.5 ns, are shown in Fig. \ref{MSPIP}. The profiles 
are scatter-broadened at 325 MHz as is evident from these figures. 
The use of incoherent dedispersion, even with 1024 channels, for 
these observations typically leaves a dispersion smear of 3.8 $\mu s~DM^{-1}$, 
which affects true scatter-broadening measurements for these 
pulsars (typical scatter-broadening tails of the order of 
0.3 ms for PSR B1937+21). For low DM MSPs, generally used in 
pulsar timing array experiments, the high time resolution 
observations, made possible by PONDER, provide low post-fit residual 
times-of-arrival.

\paragraph{} Sometimes, the dispersion smear due to incoherent dedispersion 
can hide pulse components or lead to a large uncertainty in 
estimating pulse separation or ratio of pulse components.
Fig. \ref{MSPIP} shows the average profile of 
PSR B0531$+$21, obtained with both incoherent and 
coherent dedispersion. The difference between these two profiles 
is also shown, which clearly indicates the result of dispersion smear. 
The scatter-broadening in this pulsar varies with time and sometimes 
can completely hide the precursor component (the component 
before the stronger of the  three components) at this frequency 
as the main pulse and the precursor merge. 
Such observations with PONDER are useful to study the profile evolution 
of short period radio pulsars at lower frequencies.

\begin{figure}
\includegraphics[width=1\textwidth]{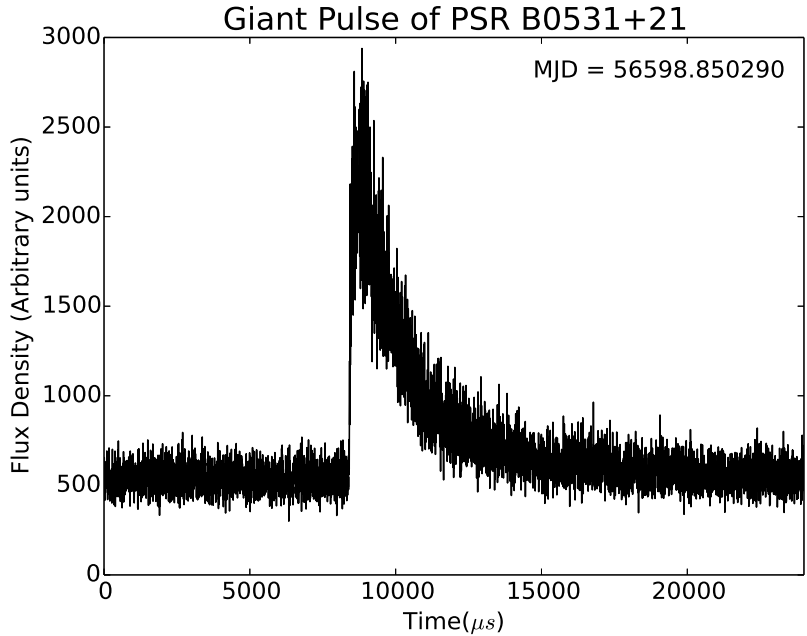}\centering
\caption{Plot of the Giant Pulse of PSR B0531+21 observed using PONDER 
at MJD 56598.850290. The flux density in arbitrary units is shown as a 
function of time. The time resolution is 1 $\mu s$. The scatter-broadening 
tail is clearly visible.}
\label{GP}
\end{figure}

\begin{figure}
\includegraphics[width=1\textwidth]{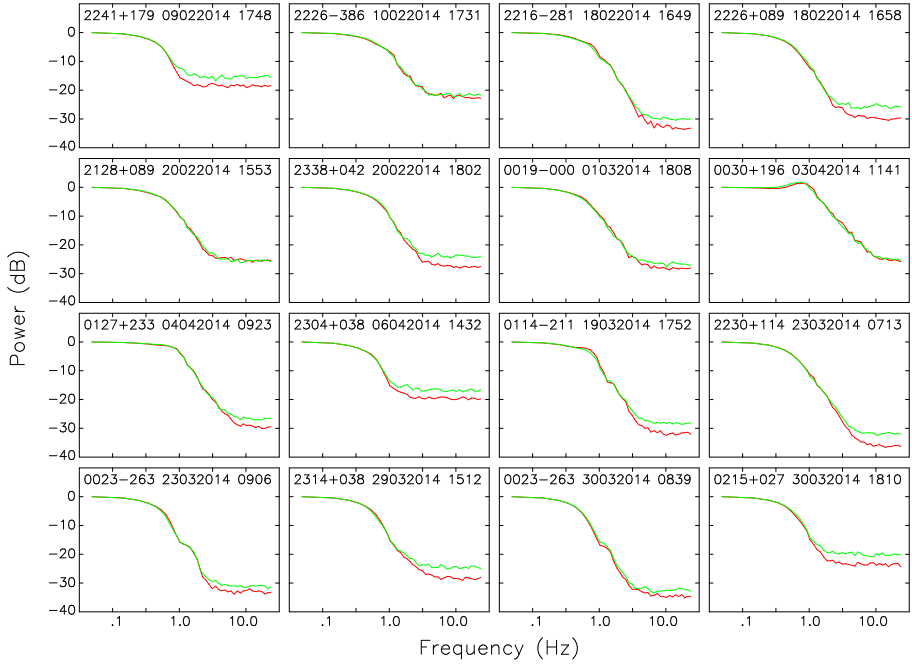}\centering
\caption{This figure shows spectra obtained from the old 4-MHz system (green color) and new PSR
system (red color). The top legend on each small plot gives the source name (B1950),
date and time of observation.}
\label{ips_psm}
\end{figure}

\paragraph{}Another area of investigations, where PONDER will be 
very useful, is  the
study of giant pulse (GP) emission and micro-structure. GPs are 
narrow pulses, with intensity several order of magnitude higher 
than the mean intensity, exhibited by a small number of pulsars 
\citep{lcu+95,kt00,jr03,jkl+04}.  
GP emission is as yet not well understood. A GP in PSR B0531+21 
observed using PONDER is shown in Fig. \ref{GP} with a time resolution of 1 
$\mu$s.
As GPs are almost impulse like, pulsars 
with scatter broadened GPs can be useful in estimating 
the true scatter-broadening time scale in these line of sight. 

\paragraph{}Lastly, PONDER has been extensively tested during IPS observations. 
Simultaneous IPS measurements on a large number of scintillating radio
sources using both PONDER and the old conventional 4-MHz
correlated-beam system at the ORT were carried out. These 
observations covered a 
period of about two months, during February and April 2014. For 
observations on a radio source, the scintillation time series from the
PONDER was integrated to 20 ms sampling to match the
conventional system observing data rate. The time series from the
above two systems have been processed using an identical analysis
procedure to yield the temporal power spectra, covering a temporal 
frequency range of 0 to 25 Hz. In Fig. \ref{ips_psm}, some of
the sample spectra obtained from PONDER (red color) and conventional
system (green color) are displayed. In this figure, the scintillating
power (in dB) is plotted as a function of logarithm of temporal
frequency (Hz). In each spectral plot, the radio source name (in B1950
format), date and time of observation are shown at the top. It is
evidently clear that the PONDER measurements reproduced each and every
feature observed in the old conventional system. In particular, 
the spectra, obtained with PONDER, reproduces specific 
features in the declining part of the spectra, seen in the 
spectra obtained by the conventional system 
for 0030+196, 0114-211 and 0023-263. 
This essentially validates the IPS observing mode of 
PONDER.

\begin{figure}
\includegraphics[width=1\textwidth]{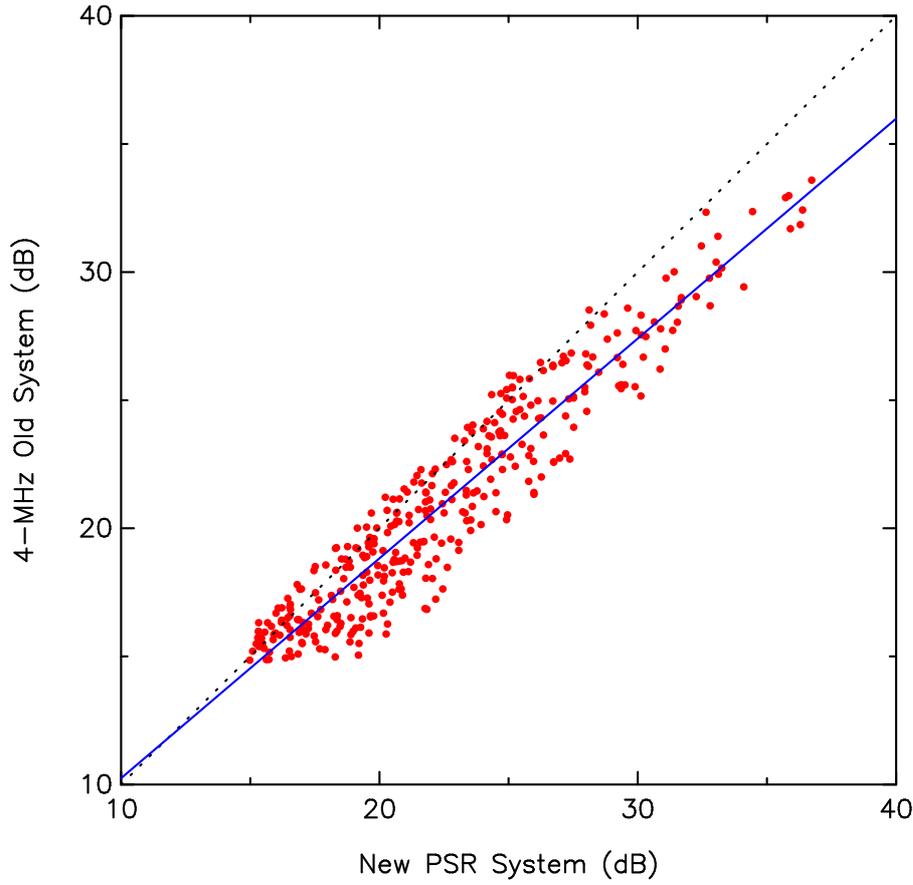}\centering
\caption{A comparison of the spectral power observed with the old 
4-MHz system and the PONDER. The x and y-axis scales are in
dB. The dotted line is the one-to-one correlation line. The solid line 
is the best fit to the data points. On the whole the PONDER  
gives higher S/N by ~3 dB. One can see higher scatter towards low 
S/N side, implying that the points lying above the dotted line 
in this region are probably affected by higher RFI
included in the wide bandwidth of the PONDER.}
\label{ips_corr}
\end{figure}

\paragraph{}  It is interesting to note that for most of the spectra, 
the PONDER measurements provide an excess signal around 1 Hz  
frequency range of the spectra. In other words, the noise level of 
the PONDER system is lower than the old system. This is consistent 
with the increase in bandwidth from the old system of 4-MHz to a 
16-MHz PONDER system. The increase in power spectrum signal is 
evident in Fig. \ref{ips_corr}, which shows the correlation plot of 
power spectrum signal observed between the old conventional 4-MHz 
system and the new 16-MHz PONDER system. This plot includes spectral 
signals of 364 simultaneous IPS observations made between the old and 
new systems. The dotted line indicates the one-to-one correlation 
line between two data sets. The continuous line is the best
straight-line fit to the observations. On the average
PONDER tends to give a 3dB increase in spectral power compared to the
old system, which is in agreement with the increase of bandwidth by
about 4 times. However, since the PONDER is a wide band system, it can
include more RFI signals than that of the old 4-MHz system and this 
could be the likely reason for the larger scatter in this figure 
for the lower signal-to-noise ratio spectra (near 20 dB) observed with 
the PONDER. Nevertheless, the PONDER system gives an
additional power at the high frequency portion of the spectrum, which
is an essential requirement to get the source size information from
the IPS temporal spectrum as well as inner-scale (i.e., cut-off scale)
size of the solar wind turbulence \citep{Mano3}. Thus, PONDER will be very 
useful for the understanding of the smallest scale in solar wind turbulence. 

\section{Summary and Future plans}
\label{summary}

\paragraph{}
A new real-time backend, PONDER, designed to operate with the legacy system of the ORT
has been described in this paper. PONDER uses current state of the art computing
hardware, a GPU board and a large disk storage to support high time resolution real-time
pulsar data by employing coherent dedispersion over a bandpass of 16 MHz. Moreover,
PONDER can be operated in a variety of observing modes using a GUI. For example, it can
essentially support different modes of pulsar observations and each mode leads to
standard reduced data products in the form of integrated pulsar profiles and dedispersed
time series, which allow a faster turn-around time from observations to scientific
results. There is ample scope for getting additional data products in future. In the case of
IPS observations, PONDER has demonstrated the improved sensitivity of the
fluctuation spectrum at the high-frequency portion of the spectrum, which is important in
getting some of the crucial solar-wind parameters. The IPS mode has also enabled the
availability of correlated time series and fluctuation spectrum products with high time
and frequency resolution in real time. Additionally, the capabilities of PONDER
illustrated by the pulsar and IPS modes can be extended to use for a variety
of other astrophysical studies possible using the high sensitivity of the ORT.

In the near future, it is proposed to add a dynamic-spectrum mode for pulsar
observations to obtain online dynamic spectra as a standard data product for the studies
of ISM using pulsars as the probe. Another enhancement will be an automated pipeline
using the gated pulsar observations planned in the dynamic spectra mode, to detect nulls
and generate estimates of nulling fractions of the pulsars. Lastly, we also plan to add
a copy of the backend to other beams of the legacy system, with additional capability of
automatic detection of transients, such as fast radio bursts \citep{tsb+13}.

\begin{acknowledgements}
We would like to thank all the staff members of the Radio Astronomy
Center for helping us in various stages of the development and testing
of the receiver. In particular, we acknowledge the help from D.
Nandagopal, K. Kalyanasundaram, Arun E Varghese, Amit Kumar Mittal \&
G.V.S Girish for helping us with various stages of the PONDER
development. We also thank all the operators of the ORT who have
assisted during the observations mentioned in the paper mainly V.
Magesh, R. Chandrasekhar, P. Praveen \& Nevile Jude. The ORT is
operated and maintained at the Radio Astronomy Centre by the National
Centre for Radio Astrophysics of the Tata Institute of Fundamental
Research. We are grateful to the anonymous referees for their 
careful and critical reading of the manuscript and useful comments.  
\end{acknowledgements}

\bibliographystyle{spbasic} 
\bibliography{journals_apj,psrrefs,modrefs,mybib,paperRef,crossrefs}

\begin{thebibliography}{35}
\providecommand{\natexlab}[1]{#1}
\providecommand{\url}[1]{{#1}}
\providecommand{\urlprefix}{URL }
\expandafter\ifx\csname urlstyle\endcsname\relax
  \providecommand{\doi}[1]{DOI~\discretionary{}{}{}#1}\else
  \providecommand{\doi}{DOI~\discretionary{}{}{}\begingroup
  \urlstyle{rm}\Url}\fi
\providecommand{\eprint}[2][]{\url{#2}}

\bibitem[{{Demorest} et~al(2013){Demorest}, {Ferdman}, {Gonzalez}, {Nice},
  {Ransom}, {Stairs}, {Arzoumanian}, {Brazier}, {Burke-Spolaor}, {Chamberlin},
  {Cordes}, {Ellis}, {Finn}, {Freire}, {Giampanis}, {Jenet}, {Kaspi}, {Lazio},
  {Lommen}, {McLaughlin}, {Palliyaguru}, {Perrodin}, {Shannon}, {Siemens},
  {Stinebring}, {Swiggum}, and {Zhu}}]{dfg+13}
{Demorest} PB, {Ferdman} RD, {Gonzalez} ME, {Nice} D, {Ransom} S, {Stairs} IH,
  {Arzoumanian} Z, {Brazier} A, {Burke-Spolaor} S, {Chamberlin} SJ, {Cordes}
  JM, {Ellis} J, {Finn} LS, {Freire} P, {Giampanis} S, {Jenet} F, {Kaspi} VM,
  {Lazio} J, {Lommen} AN, {McLaughlin} M, {Palliyaguru} N, {Perrodin} D,
  {Shannon} RM, {Siemens} X, {Stinebring} D, {Swiggum} J, {Zhu} WW (2013)
  {Limits on the Stochastic Gravitational Wave Background from the North
  American Nanohertz Observatory for Gravitational Waves}. ApJ 762:94,
  \doi{10.1088/0004-637X/762/2/94}, \eprint{1201.6641}

\bibitem[{{DuPlain} et~al(2008){DuPlain}, {Ransom}, {Demorest}, {Brandt},
  {Ford}, and {Shelton}}]{dsd+08}
{DuPlain} R, {Ransom} S, {Demorest} P, {Brandt} P, {Ford} J, {Shelton} AL
  (2008) {Launching GUPPI: the Green Bank Ultimate Pulsar Processing
  Instrument}. In: Society of Photo-Optical Instrumentation Engineers (SPIE)
  Conference Series, Society of Photo-Optical Instrumentation Engineers (SPIE)
  Conference Series, vol 7019, \doi{10.1117/12.790003}

\bibitem[{Foster and Backer(1990)}]{fb90}
Foster RS, Backer DC (1990) Constructing a pulsar timing array. ApJ 361:300

\bibitem[{{Hankins}(1971)}]{Hankins}
{Hankins} TH (1971) {Microsecond Intensity Variations in the Radio Emissions
  from CP 0950}. \apj 169:487, \doi{10.1086/151164}

\bibitem[{{Hankins} and {Rickett}(1975{\natexlab{a}})}]{hr75}
{Hankins} TH, {Rickett} BJ (1975{\natexlab{a}}) Pulsar signal processing. In:
  Methods in Computational Physics Volume 14 --- Radio Astronomy, Academic
  Press, New York, pp 55--129

\bibitem[{{Hankins} and {Rickett}(1975{\natexlab{b}})}]{Hankins1}
{Hankins} TH, {Rickett} BJ (1975{\natexlab{b}}) {Pulsar signal processing.}
  Methods in Computational Physics 14:55--129

\bibitem[{{Hankins} et~al(2003){Hankins}, {Kern}, {Weatherall}, and
  {Eilek}}]{hkwe03}
{Hankins} TH, {Kern} JS, {Weatherall} JC, {Eilek} JA (2003) {Nanosecond radio
  bursts from strong plasma turbulence in the Crab pulsar}. Nature 422:141--143

\bibitem[{{Hobbs} et~al(2006){Hobbs}, {Edwards}, and {Manchester}}]{tempo2}
{Hobbs} GB, {Edwards} RT, {Manchester} RN (2006) {TEMPO2, a new pulsar-timing
  package - I. An overview}. \mnras 369:655--672,
  \doi{10.1111/j.1365-2966.2006.10302.x}, \eprint{astro-ph/0603381}

\bibitem[{Jenet et~al(1998)Jenet, Anderson, Kaspi, Prince, and Unwin}]{jak+98}
Jenet F, Anderson S, Kaspi V, Prince T, Unwin S (1998) {Radio pulse properties
  of the millisecond pulsar PSR J0437$-$4715. I. Observations at 20
  centimeters}. ApJ 498:365--372

\bibitem[{Johnston and Romani(2002)}]{jr02}
Johnston S, Romani R (2002) {A search for giant pulses in Vela-like pulsars}.
  MNRAS 332:109--115

\bibitem[{Johnston and Romani(2003)}]{jr03}
Johnston S, Romani R (2003) Giant pulses from psr b0540-69 in the large
  magellanic cloud. ApJ 590:L95--L98

\bibitem[{{Joshi} and {Ramakrishna}(2006)}]{jr06}
{Joshi} BC, {Ramakrishna} S (2006) {A software baseband receiver for pulsar
  astronomy at GMRT}. Bulletin of the Astronomical Society of India 34:401,
  \eprint{astro-ph/0611331}

\bibitem[{{Joshi} et~al(2003){Joshi}, {Lyne}, {Kramer}, {Lorimer}, {Jordan},
  {Holloway}, {Ikin}, and {Stairs}}]{jlk+03}
{Joshi} BC, {Lyne} AG, {Kramer} M, {Lorimer} DR, {Jordan} C, {Holloway} A,
  {Ikin} T, {Stairs} IH (2003) {Coherent On-line Baseband Receiver for
  Astronomy}. In: Bailes M, Nice DJ, Thorsett S (eds) Radio Pulsars,
  Astronomical Society of the Pacific, San Francisco, p 321

\bibitem[{{Joshi} et~al(2004){Joshi}, {Kramer}, {Lyne}, {McLaughlin}, and
  {Stairs}}]{jkl+04}
{Joshi} BC, {Kramer} M, {Lyne} AG, {McLaughlin} MA, {Stairs} IH (2004) {Giant
  Pulses in Millisecond Pulsars}. In: Camilo F, Gaensler BM (eds) IAU
  Symposium, p 319

\bibitem[{{Karuppusamy} et~al(2008){Karuppusamy}, {Stappers}, and {van
  Straten}}]{ksv08}
{Karuppusamy} R, {Stappers} B, {van Straten} W (2008) {PuMa-II: A Wide Band
  Pulsar Machine for the Westerbork Synthesis Radio Telescope}. Proc Astr Soc
  Aust 120:191--202, \doi{10.1086/528699}

\bibitem[{Kinkhabwala and Thorsett(2000)}]{kt00}
Kinkhabwala A, Thorsett SE (2000) Multifrequency observations of giant radio
  pulses from the millisecond pulsar~{B}1937+21. ApJ 535:365--372

\bibitem[{{Kramer}(1998)}]{kra98}
{Kramer} M (1998) Determination of the geometry of the {PSR~B1913+16} system by
  geodetic precession. ApJ 509:856--860

\bibitem[{{Kramer} et~al(2002){Kramer}, {Johnston}, and {van Straten}}]{kjv02}
{Kramer} M, {Johnston} S, {van Straten} W (2002) {High-resolution single-pulse
  studies of the Vela pulsar}. MNRAS 334:523--532

\bibitem[{{Kramer} et~al(2006){Kramer}, {Stairs}, {Manchester}, {McLaughlin},
  {Lyne}, {Ferdman}, {Burgay}, {Lorimer}, {Possenti}, {D'Amico}, {Sarkissian},
  {Hobbs}, {Reynolds}, {Freire}, and {Camilo}}]{ksm+06}
{Kramer} M, {Stairs} IH, {Manchester} RN, {McLaughlin} MA, {Lyne} AG, {Ferdman}
  RD, {Burgay} M, {Lorimer} DR, {Possenti} A, {D'Amico} N, {Sarkissian} JM,
  {Hobbs} GB, {Reynolds} JE, {Freire} PCC, {Camilo} F (2006) {Tests of General
  Relativity from Timing the Double Pulsar}. Science 314:97--102,
  \doi{10.1126/science.1132305}

\bibitem[{{Krishnakumar} et~al(2015){Krishnakumar}, {Mitra}, {Naidu}, {Joshi},
  and {Manoharan}}]{k15}
{Krishnakumar} MA, {Mitra} D, {Naidu} A, {Joshi} BC, {Manoharan} PK (2015)
  {Scatter broadening measurements of 124 pulsars at 327 MHz}. ArXiv e-prints
  \eprint{1501.05401}

\bibitem[{{Lorimer}(2001)}]{lor01b}
{Lorimer} DR (2001) {SIGPROC-v1.0: (Pulsar) Signal Processing Programs},
  {Arecibo Technical Memo No.~2001--01}

\bibitem[{Lundgren et~al(1995)Lundgren, Cordes, Ulmer, Matz, Lomatch, Foster,
  and Hankins}]{lcu+95}
Lundgren SC, Cordes JM, Ulmer M, Matz SM, Lomatch S, Foster RS, Hankins T
  (1995) Giant pulses from the crab pulsar: A joing radio and gamma-ray study.
  ApJ 453:433--445

\bibitem[{Lyne et~al(2004)Lyne, Burgay, Kramer, Possenti, Manchester, Camilo,
  McLaughlin, Lorimer, D'Amico, Joshi, Reynolds, and Freire}]{lbk+04}
Lyne AG, Burgay M, Kramer M, Possenti A, Manchester RN, Camilo F, McLaughlin
  MA, Lorimer DR, D'Amico N, Joshi BC, Reynolds J, Freire PCC (2004) A
  double-pulsar system: {A} rare laboratory for relativistic gravity and plasma
  physics. Science 303:1153--1157

\bibitem[{{Manchester} et~al(2013){Manchester}, {Hobbs}, {Bailes}, {Coles},
  {van Straten}, {Keith}, {Shannon}, {Bhat}, {Brown}, {Burke-Spolaor},
  {Champion}, {Chaudhary}, {Edwards}, {Hampson}, {Hotan}, {Jameson}, {Jenet},
  {Kesteven}, {Khoo}, {Kocz}, {Maciesiak}, {Oslowski}, {Ravi}, {Reynolds},
  {Sarkissian}, {Verbiest}, {Wen}, {Wilson}, {Yardley}, {Yan}, and
  {You}}]{mhb+13}
{Manchester} RN, {Hobbs} G, {Bailes} M, {Coles} WA, {van Straten} W, {Keith}
  MJ, {Shannon} RM, {Bhat} NDR, {Brown} A, {Burke-Spolaor} SG, {Champion} DJ,
  {Chaudhary} A, {Edwards} RT, {Hampson} G, {Hotan} AW, {Jameson} A, {Jenet}
  FA, {Kesteven} MJ, {Khoo} J, {Kocz} J, {Maciesiak} K, {Oslowski} S, {Ravi} V,
  {Reynolds} JR, {Sarkissian} JM, {Verbiest} JPW, {Wen} ZL, {Wilson} WE,
  {Yardley} D, {Yan} WM, {You} XP (2013) {The Parkes Pulsar Timing Array
  Project}. Proc Astr Soc Aust 30:e017, \doi{10.1017/pasa.2012.017},
  \eprint{1210.6130}

\bibitem[{{Manoharan}(2012)}]{Mano1}
{Manoharan} PK (2012) {Three-dimensional Evolution of Solar Wind during Solar
  Cycles 22-24}. \apj 751:128, \doi{10.1088/0004-637X/751/2/128},
  \eprint{1203.6715}

\bibitem[{{Manoharan} et~al(2000){Manoharan}, {Kojima}, {Gopalswamy}, {Kondo},
  and {Smith}}]{Mano3}
{Manoharan} PK, {Kojima} M, {Gopalswamy} N, {Kondo} T, {Smith} Z (2000) {Radial
  Evolution and Turbulence Characteristics of a Coronal Mass Ejection}. \apj
  530:1061--1070, \doi{10.1086/308378}

\bibitem[{{Manoharan} et~al(2001){Manoharan}, {Tokumaru}, {Pick},
  {Subramanian}, {Ipavich}, {Schenk}, {Kaiser}, {Lepping}, and
  {Vourlidas}}]{Mano2}
{Manoharan} PK, {Tokumaru} M, {Pick} M, {Subramanian} P, {Ipavich} FM, {Schenk}
  K, {Kaiser} ML, {Lepping} RP, {Vourlidas} A (2001) {Coronal Mass Ejection of
  2000 July 14 Flare Event: Imaging from Near-Sun to Earth Environment}. \apj
  559:1180--1189, \doi{10.1086/322332}

\bibitem[{Popov et~al(2002)Popov, Bartel, Cannon, Novikov, Kondratiev, and
  Altunin}]{pbc+02}
Popov MV, Bartel N, Cannon WH, Novikov AY, Kondratiev VI, Altunin VI (2002)
  Pulsar microstructure and its quasi-periodicities with the s2 vlbi system at
  a resolution of 62.5 nanoseconds. A\&A pp 171--187

\bibitem[{{Romani} et~al(1986){Romani}, {Narayan}, and {Blandford}}]{Romani}
{Romani} RW, {Narayan} R, {Blandford} R (1986) {Refractive effects in pulsar
  scintillation}. \mnras 220:19--49

\bibitem[{Sallmen and Backer(1995)}]{sb95}
Sallmen S, Backer DC (1995) Single pulse statistics for {PSR}~1534+12 and
  {PSR}~1937+21. In: Fruchter AS, Tavani M, Backer DC (eds) Millisecond
  Pulsars: A Decade of Surprise, Astron.\ Soc.\ Pac.\ Conf.\ Ser.\ Vol.\ 72, pp
  340--342

\bibitem[{van Straten et~al(2001)van Straten, Bailes, Britton, Kulkarni,
  Anderson, Manchester, and Sarkissian}]{vbb+01}
van Straten W, Bailes M, Britton M, Kulkarni SR, Anderson SB, Manchester RN,
  Sarkissian J (2001) A test of general relativity from the three-dimensional
  orbital geometry of a binary pulsar. Nature 412:158--160

\bibitem[{{Swarup} et~al(1971){Swarup}, {Sarma}, {Joshi}, {Kapahi}, {Bagri},
  {Damle}, {Ananthakrishnan}, {Balasubramanian}, {Bhave}, and
  {Sinha}}]{swarup1}
{Swarup} G, {Sarma} NVG, {Joshi} MN, {Kapahi} VK, {Bagri} DS, {Damle} SH,
  {Ananthakrishnan} S, {Balasubramanian} V, {Bhave} SS, {Sinha} RP (1971)
  {Large Steerable Radio Telescope at Ootacamund, India}. Nature Physical
  Science 230:185--188, \doi{10.1038/physci230185a0}

\bibitem[{Taylor and Weisberg(1989)}]{tw89}
Taylor JH, Weisberg JM (1989) Further experimental tests of relativistic
  gravity using the binary pulsar {PSR}\,1913+16. ApJ 345:434--450

\bibitem[{{Thornton} et~al(2013){Thornton}, {Stappers}, {Bailes}, {Barsdell},
  {Bates}, {Bhat}, {Burgay}, {Burke-Spolaor}, {Champion}, {Coster}, {D'Amico},
  {Jameson}, {Johnston}, {Keith}, {Kramer}, {Levin}, {Milia}, {Ng}, {Possenti},
  and {van Straten}}]{tsb+13}
{Thornton} D, {Stappers} B, {Bailes} M, {Barsdell} B, {Bates} S, {Bhat} NDR,
  {Burgay} M, {Burke-Spolaor} S, {Champion} DJ, {Coster} P, {D'Amico} N,
  {Jameson} A, {Johnston} S, {Keith} M, {Kramer} M, {Levin} L, {Milia} S, {Ng}
  C, {Possenti} A, {van Straten} W (2013) {A Population of Fast Radio Bursts at
  Cosmological Distances}. Science 341:53--56, \doi{10.1126/science.1236789},
  \eprint{1307.1628}

\bibitem[{{Weisberg} and {Taylor}(2002)}]{wt02}
{Weisberg} JM, {Taylor} JH (2002) {General Relativistic Geodetic Spin
  Precession in Binary Pulsar B1913+16: Mapping the Emission Beam in Two
  Dimensions}. ApJ 576:942--949

\end{thebibliography}

\end{document}